\documentclass[aps,pra,showpacs,superscriptaddress,groupedaddress]{revtex4}  
\usepackage{graphicx}  
\usepackage{dcolumn}   
\usepackage{bm}        
\usepackage{amssymb}   
\usepackage{amsmath} 
\makeatletter 
\def\maketag@@@#1{\hbox{\m@th\normalfont\normalsize#1}}
\makeatother 
\usepackage[toc,page]{appendix} 

\usepackage{color}
\usepackage{graphicx}
\usepackage{amssymb}
\usepackage{amsmath}
\usepackage{amsfonts}
\usepackage[mathscr]{eucal}
\usepackage{bbold}
\usepackage{color}
\usepackage{tikz}
\usetikzlibrary{matrix,arrows,decorations.pathmorphing}
\usepackage{framed}

\newcommand{\ket}[1]{\left| #1 \right\rangle}

\newcommand{\braket}[2]{\left\langle #1 | #2 \right\rangle}
\newcommand{\proj}[1]{| #1\rangle\!\langle #1 |}
\newcommand{\ketbra}[2]{| #1\rangle\!\langle #2 |}

\newcommand{\Tr}{\mathrm{Tr}}

\renewcommand{\Re}{\mathrm{Re}}

\newcommand{\T}{\textsf{\scriptsize T}}

\newcommand{\inv}{{\,\text{-}\hspace{-1pt}1}}
\newcommand{\ad}{\mathrm{ad}}
\newcommand{\Ad}{\mathrm{Ad}}
\newcommand{\Z}{\mathcal{Z}}
\newcommand{\K}{\mathcal{K}}

\newcommand{\D}{\mathcal{D}}

\newcommand{\ZZ}{\mathbb Z}
\newcommand{\R}{\mathbb R}
\newcommand{\C}{\mathbb C}
\newcommand{\U}{\mathrm U}
\newcommand{\SO}{\mathrm{SO}}
\newcommand{\SL}{\mathrm{SL}}
\newcommand{\SU}{\mathrm{SU}}
\newcommand{\Sp}{\mathrm{Sp}}
\renewcommand{\u}{\mathfrak u}
\newcommand{\su}{\mathfrak{su}}
\renewcommand{\sl}{\mathfrak{sl}}
\newcommand{\so}{\mathfrak{so}}
\renewcommand{\sp}{\mathfrak{sp}}
\newcommand{\Hb}{\mathcal H}

\newcommand{\g}{\mathfrak g}
\newcommand{\h}{\mathfrak h}

\newcommand{\csch}{\mathrm{csch}}

\def\dbar{{\mathchar'26\mkern-12mu d}}

\renewenvironment{framed}[1][\hsize]
{\MakeFramed{\hsize#1\advance\hsize-\width \FrameRestore}}%
{\endMakeFramed}

\begin{document}

\title{How to implement a generalized-coherent-state POVM\\ via nonadaptive continuous isotropic measurement}

\author{Christopher S. Jackson}
\affiliation{Center for Quantum Information and Control (CQuIC) \\ 
Department of Physics and Astronomy, University of New Mexico, Albuquerque, NM 87131, USA}

\date{\today}
\begin{abstract}
In a recent Letter~[PRL \textbf{121}, 130404 (2018)], it was announced that the spin-coherent-state POVM can be implemented via a nonadaptive continuous isotropic measurement.
In this article, the mathematical concepts used to prove this are explained in greater depth.
Also provided is the more general result of how to implement a generalized-coherent-state POVM for any finite-dimensional unitary representation of a Lie group.
\end{abstract}

\maketitle

\section{Introduction}

Originially coined by Perelomov, the term \emph{generalized coherent state}~\cite{perelomov1972} has found various definitions of varying generality.
For our purposes, we define generalized coherent states in the context of Hilbert spaces that carry a unitary irreducible representation (unirrep) of a \textbf{compact} Lie group.
In particular, we define generalized coherent states (GCSs) as states in the orbit of \textbf{highest weight}.
Although the original Perelomov prescription does not specify which fiducial state from which to generate an orbit, it is important to point out that it is this orbit of highest weight that is the closest analogue to the most utilized of coherent states, those from the legacy of Roy Glauber, who coined the term \emph{coherent state} and demonstrated the utility of these states in quantum optics.

A quick explanation of these analogous properties of highest weight GCSs and Glauber coherent states is in order.
When a separable Hilbert space carries a unirrep of a Lie group, all Hamiltonians over such a Hilbert space can be expressed as polynomials in the representations of the corresponding Lie algebra.
Highest weight GCSs have the easiest-to-calculate expectation values precisely because they are annihilated by all raising operators up to a unitary generated by a linear Hamiltonian|that is, a Hamiltonian representing an element of the Lie algebra.
What is distinct between highest weight GCSs and Glauber coherent states is that the normalizer of linear Hamiltonians is identical to the original Lie group.
For this reason, GCSs are sometimes also referred to as ``Gaussian'' states.

By Schur's lemma, GCSs form a continuous resolution of the identity and thus define a POVM as well as a phase-space correspondence.
Therefore, GCSs provide an entire paradigm for conceptualizing quantum systems.
Despite these relatively well-known facts, the measurements and phase spaces of continuous GCSs are not nearly as utilized in physics as the quantum-optics coherent states of Glauber.
Perhaps the reason for this is precisely the absence of a known practical implementation of the GCS measurement.

The simplest case of highest-weight GCSs are the spin-coherent states carrying a spin-$j$ unirrep of the rotation group $\SO(3)$.
An important instance of these spin-coherent states is in $2j$ copies of a qubit.
In this context, it has been known~\cite{Massar1995} for some time that the spin-coherent-state POVM represents a measurement which could estimate an unknown qubit with the largest possible average fidelity.
From this, two conversations arose in an attempt to discover a practical implementation of the spin-coherent-state measurement which ultimately failed.
The first of these conversations worked off the idea that for $N=2j$ copies, one can replace the spin-coherent-state measurement with a discrete POVM consisting of finitely many outcomes that compose a minimal spherical $N$-design which could be implemented via Neumark extension, however such measurements are not amenable to general spins~\cite{PhysRevLett.80.1571, PhysRevLett.81.1351, BRU1999249, PhysRevA.61.022113}.
The second of these conversations developed the idea that since for Glauber-coherent states heterodyne measurements are a single-shot implementation of uniform homodyne measurements, perhaps spin-coherent-state measurements have an analogous ``single-shot'' implementation via isotropic spin-component measurements~\cite{peres2006quantum,D'Ariano2001,D'Ariano2002}.

With the recent discovery of an implementation of the spin-coherent-state measurement via nonadaptive continuous isotropic measurement, it becomes apparent that there is one dominant property of GCSs that must be appreciated.
This property is that GCSs are associated with manifolds of constant positive curvature.
This constant curvature is both what prevents a single-shot implementation from existing while also making a non-adaptive weak measurement work.
The construction and proof of this non-adaptive protocol is the result of this article.
In summary, we show that GCSs and GCS measurements are also ``Gaussian'' in the sense that they are the universal limits of composing independent Kraus operators which represent a (semisimple) Lie group.

It is the author's personal hope that these results, by connecting GCSs to measurements, give rise to new perspectives, both theoretical and experimental, on the subtleties of curved phase space and curved measurement.

The layout of this paper is as follows.
Section \ref{formalspin} is a formal overview of the group-theoretic aspects of quantum spin.
Section \ref{SCSPOVM} reviews the nonadaptive isotropic continuous measurement of \cite{shojaee2018optimal}.
Section \ref{GCSPOVM} introduces the generalized-coherent-state POVM and demonstrates analogous results for the isotropic continuous measurement.

\section{The group theoretic formalism of rotation and quantum spin}\label{formalspin}

This section is a formal overview of the theory of rotation, quantum spin, spin-coherent states, and the spin-coherent-state POVM.
The primary purpose of this section is to allow the reader, who might be much more familiar with spin than most other Lie group representations, to make clear analogies with the generalized-coherent states and their POVM.
Emphasis is made on group-theoretic techniques and their associated perspectives.

\vspace{20pt}

Define the special linear groups over a continuous field $\mathbb{K}$ (either $\R$ or $\C$)
\begin{equation}
\SL_\mathbb{K}(n) = \Big\{ n\!\times\!n \text{ matrices, } M\text{, with entries in } \mathbb{K} : \det M = 1 \Big\},
\end{equation}
the special unitary group
\begin{equation}
\SU(n) = \Big\{ U \in \SL_\C(n): U^\dag U = 1 \Big\},
\end{equation}
the line $\R$ under addition, and the circle
\begin{equation}
\U(1) = \big\{z \in \C : z^*z=1\big\}
\end{equation}
under multiplication.
The circle $\U(1)$ has other unitary irreducible representations (unirreps,) all of which are one-dimensional.
The unirreps of the circle can be enumerated by powers $m$
\begin{equation}\label{powers}
U(e^{-i\phi}) = e^{-im\phi}
\end{equation}
which are integer for nonprojective unirreps and a fraction for projective unirreps.
The universal covering group of $\U(1)$ under multiplication is isomorphic to $\R$ under addition with fiber isomorphic to $\ZZ$.
If a Hilbert space $\Hb$ has dimension $n$, it is convenient to use the notation
\begin{equation}
\SU(\Hb) \cong \SU(n)
\hspace{10pt}
\text{,}
\hspace{50pt}
\SL_\C(\Hb) \cong \SL_\C(n)
\hspace{10pt}
\text{, }
\hspace{30pt}
\text{etc.}
\end{equation}
Under matrix multiplication, each of these groups is a Lie group which means that every group element is the exponential of a so-called infinitesimal generator.
Under commutators, the infinitesimal generators define Lie algebras
\begin{equation}
\sl_\mathbb{K}(3) = \Big\{ 3\!\times\!3 \text{ matrices, } X\text{, with entries in } \mathbb{K} : \Tr X = 0 \Big\},
\end{equation}
\begin{equation}
\su(n) = \Big\{ iH \in \sl_\C(n): (iH)^\dag = -iH \Big\},
\end{equation}
and
\begin{equation}
\u(1) = \{i\phi: \phi \in \R\}.
\end{equation}

\subsection{The rotation group and spin}

Spin is an order parameter defined by its transformation properties under rotation.
The set of rotations in 3 dimensions under composition form a group, isomorphic to the special orthogonal matrix group
\begin{equation}\label{defininG}
\SO(3) = \Big\{ R \in \SL_\R(3) : R^\T R = 1 \Big\}.
\end{equation}
$\SO(3)$ is also a Lie group, with Lie algebra
\begin{equation}\label{defining}
\so(3) = \Big\{ A \in \sl_\R(3) : A^\T = -A \Big\}.
\end{equation}
A standard basis for $\so(3)$ is given by the three generators $L_k$ with matrix elements
\begin{equation}
(L_k)_{ij} = \epsilon_{ijk}
\end{equation}
where $\epsilon_{ijk}$ is the antisymmetric Levi-Civita symbol.
Equations \ref{defininG} and \ref{defining} are called the \emph{defining representations} of the Lie group $\SO(3)$ and Lie algebra $\so(3)$.
This group has other inequivalent unitary irreducible representations (unirrep) carried by Hilbert spaces of every dimension.
Infinitesimally, we have unirreps defined by antiHermitian operators
\begin{equation}
-iJ:\so(3)\longrightarrow\su(2j+1)
\end{equation}
where $j$ can be a half-integer or integer.
Representations are irreducible if they cannot be simultaneously block-diagonalized.
Define
\begin{equation}
J_k = J(L_k)
\end{equation}
and denote general elements of the representation by
\begin{equation}
\psi\, \hat n \cdot \vec J = \psi\, n^k J_k.
\end{equation}
These parameters correspond to the angle, $\psi$, and axis, $\hat n$, of a rotation.
These representations of the Lie algebra exponentiate to representations of the rotation group, except that for even dimensions these representations of $\SO(3)$ become projective because
\begin{equation}
e^{-i2\pi \hat n \cdot \vec J} = (-1)^{2j}.
\end{equation}
Projective representations are handled by the fact that the rotation group can be ``extended'' by a finite abelian group to another Lie group which has no projective representations, the so called universal covering group.  The universal cover for 3-dimensional rotation is isomorphic to the unitary group of qubits, $\SU(2)$ with fiber isomorphic to $\ZZ_2$.
The defining representation of $\SU(2)$ is said to be a \emph{fundamental represenation} of $\SO(3)$
and for the two-dimensional irrep it is standard to choose
\begin{equation}
J_k = \frac{1}{2}\sigma_k
\end{equation}
where the $\sigma_k$ are the Pauli matrices.
Thus defined are Lie group unirreps
\begin{equation}
U: \SU(2) \longrightarrow \SU(2j+1)
\end{equation}
given by
\begin{equation}
U(e^{-i\frac{\psi}{2} \hat n \cdot \vec \sigma}) = U(e^{-i\psi \hat n \cdot \vec L},\pm) = e^{-i\psi \hat n \cdot \vec J}.
\end{equation}
Conjugation of the Lie algebra by the Lie group is the same in every representation (except the trivial one) and is called the \emph{adjoint representation}
\begin{equation}
U(R)J_kU(R)^\dag = J_l{R^l}_k
\end{equation}
which is equivalent between the Lie groups $\SU(2)$ and $\SO(3)$.

\subsection{Powers and copies}

Just as the unirreps of $\U(1)$ are equivalent to fractional powers of its defining unirrep
\begin{equation}
1_1^{\otimes m} \cong 1_m,
\end{equation}
the unirreps of $\SO(3)$ are equivalent to symmetrized tensor powers of its fundamental unirrep,
\begin{equation}
2^{\otimes 2j} \cong \ldots \oplus 2j+1.
\end{equation}
Explicitly, the equivalence is made by considering representations of permutation $\pi \in S_{2j}$,
\begin{equation}
U(\pi)\ket{\psi_1}\ket{\psi_2}\cdots\ket{\psi_{2j-1}}\ket{\psi_{\pi^\inv(2j)}} = \ket{\psi_{\pi^\inv(1)}}\ket{\psi_{\pi^\inv(2)}}\cdots\ket{\psi_{\pi^\inv(2j-1)}}\ket{\psi_{\pi^\inv(2j)}}.
\end{equation}
These representations commute with the tensor product representation of $\SU(2)$
\begin{equation}
U(\pi) \left(e^{-i\frac{1}{2}\theta^k \sigma_k}\right)^{\!\otimes 2j} =  \left(e^{-i\frac{1}{2}\theta^k \sigma_k}\right)^{\!\otimes 2j} U(\pi).
\end{equation}
Projection onto the completely symmetric subspace is thus given by
\begin{equation}
\Pi = \frac{1}{(2j)!}\sum_{\pi \in S_{2j}}U(\pi)
\end{equation}
and the spin-j unirreps are isomorphic to
\begin{equation}
e^{-i \theta^k J_k} = \Pi \left(e^{-\frac{i}{2}\theta^k \sigma_k}\right)^{\!\otimes 2j}\Pi.
\end{equation}

\subsection{von Neumann measurement and quantum numbers}

Representations are important in quantum theory as the dimensions of a Hilbert space which carries a representation also carries a description of quantum measurements and their outcomes.
Observables are thus associated with the infinitesimal generators of rotation, their Hermitian counterparts are called spins, and the observables of the standard basis are called spin-components.
The number $j$ which enumerates the unirreps of $\SO(3)$ is called a quantum number.
Every spin is equivalent under conjugation by the group to a generator proportional to the spin-component $J_z$.
The spin-component $J_z$ generates a subgroup of $\SO(3)$ isomorphic to the circle $\U(1)$.
Under this subgroup, an $\SO(3)$-irreducible representation is $\U(1)$-reducible.
A unirrep of $\SO(3)$ with quantum number $j$ reduces to $2j+1$ inequivalent unirreps of $\U(1)$ with quantum numbers
\begin{equation}
m = -j, -j +1, \ldots , j-1, j.
\end{equation}
Equivalently, the $\SO(3)$ unirrep of dimension $2j+1$ is a direct sum of inequivalent $\U(1)$ unirreps
\begin{equation}
2j+1 \cong 1_{-j} \oplus 1_{-j+1} \oplus \ldots 1_{j-1} \oplus 1_j = \bigoplus_{m=-j}^j 1_m.
\end{equation}
Equivalently, the spin-component $J_z$ diagonalizes as
\begin{equation}
J_z = \sum_{m=-j}^j m\proj{m}
\end{equation}
and the generated subgroup isomorphic to the circle diagonalizes as
\begin{equation}
e^{-i\psi J_z} = \sum_{m=-j}^j e^{-im\psi}\proj{m}.
\end{equation}
The eigenvalues of $J_z$ are called \emph{weights} and their spectrum has a reflection symmetry for every unirrep $j$, instantiated as rotation about any axis in the $xy$-plane by an angle of $\pi$.
The Condon-Shortley convention is to choose for the reflection
\begin{equation}
W = e^{-i\pi J_y}
\end{equation}
so that
\begin{equation}
W J_z W^\dag = -J_z
\hspace{50pt}
\text{and}
\hspace{50pt}
W\ket{m} = i^{2m}\ket{-m}.
\end{equation}

\subsection{Weight and spin-coherent states}

The remainder of the Lie algebra of $\SO(3)$ decomposes into nondegenerate eigenvectors of $J_z$ under the adjoint representation,
\begin{equation}
\ad_{J_z}(J_\pm) = [J_z,J_\pm] = \pm J_\pm
\hspace{50pt}
\text{where}
\hspace{50pt}
J_\pm = J_x \pm i J_y.
\end{equation}
The nonzero eigenvalues of the adjoint representation are called the \emph{roots} because
\begin{equation}
\det(z-\ad_{\gamma^kJ_k}) = \det(z-\ad_{\gamma J_z}) = z(z-\gamma)(z+\gamma)
\hspace{50pt}
\text{where}
\hspace{50pt}
\gamma^2 = (\gamma^x)^2+(\gamma^y)^2+(\gamma^z)^2.
\end{equation}
The eigenvectors for the roots are referred to as ladder operators because they have the property that for a spin-$j$ unirreps,
\begin{equation}
J_\pm \ket{m} = \sqrt{j(j+1) - m(m\pm1)}\ket{m\pm1}.
\end{equation}
In particular, the state with highest weight, is uniquely defined by the property that
\begin{equation}
J_+ \ket{j} = 0.
\end{equation}
Equivalently, the state with highest weight is the unique groundstate of $-J_z$.
Spin-coherent states (SCSs) are usually defined as the states obtained by rotating the state with highest weight.
Equivalently, SCSs are the unique groundstates of Hamiltonians linear in the spins
\begin{equation}
-\gamma^k J_k \ket{j,\hat n} = - j\gamma \ket{j,\hat n}
\end{equation}
Projectively, these states define a manifold homeomorphic to a 2-sphere.
A choice in phase must be made and the Wigner convention is to define SCSs as
\begin{equation}
\ket{j,\hat n} = e^{-i\phi J_z}e^{-i\theta J_y}\ket{j}
\end{equation}
homeomorphic to
\begin{equation}
\hat n = e^{\phi L_z}\circ e^{\theta L_y} (\hat z) = (\hat x \cos\phi + \hat y \sin\phi)\sin\theta + \hat z \cos\theta.
\end{equation}

\subsection{Invariants integral and quadratic: the SCS POVM and spin-purity}

The group $\SU(2)$ has a quadratic Casimir operator,
\begin{equation}
\vec{J}^{\,2} \equiv J_x^2 + J_y^2 + J_z^2,
\end{equation}
uniquely defined up to normalization by its invariance
\begin{equation}
[J_k,\vec{J}^{\,2}] = 0.
\end{equation}
By Schur's lemma, the Casimir invariant is proportional to the identity for each unirrep with eigenvalue
\begin{equation}
\vec{J}^{\,2} = j(j+1)\boldsymbol{1}_{2j+1}
\end{equation}
for spin-$j$.
The group $\SU(2)$ also has a Haar measure, defined uniquely up to normalization by its invariance
\begin{equation}
d\mu(SR) = d\mu(R) = d\mu(RS).
\end{equation}
In Euler co\"ordinates,
\begin{equation}
R = e^{-i \frac{\phi}{2}\sigma_z}e^{-i \frac{\theta}{2}\sigma_y}e^{-i \frac{\psi}{2}\sigma_z},
\end{equation}
the Haar measure is
\begin{equation}
\int_{\SU(2)}\!\!d\mu\left(R\right)
= \int_{S^2}\!\!d\mu\big(\hat{n}(\theta,\phi)\big)\int_0^{4\pi}\frac{d\psi}{4\pi}
\end{equation}
where the measure of the 2-sphere (in spherical co\"ordinates) is
\begin{equation}
\int_{S^2}\!\!d\mu\big(\hat{n}(\theta,\phi)\big) = \int_0^{2\pi}\frac{d\phi}{2\pi}\int_{-1}^{1}\frac{d\cos\theta}{2}.
\end{equation}
Also by Schur's lemma,
\begin{equation}
\int_{\SU(2)}\!\!d\mu(R) U(R)\proj{j} U(R)^\dag
= \int_{S^2}\!\!d\mu(\hat n) \proj{j,\hat n}
= \frac{1}{2j+1}\boldsymbol{1}_{2j+1}.
\end{equation}
Thus define the SCS POVM
\begin{equation}
\big(dE_j(\hat n)\big| \equiv (2j+1) d\mu(\hat n) \proj{j,\hat n}.
\end{equation}
Define the ``\emph{spin-impurity}''~\cite{somma2018quantum, oszmaniec2014applications}
\begin{equation}\label{spinimpurity}
\bar{P}_j(E) = 1 - \sum_k \frac{(E|J_k)(J_k|E)}{j^2 (E|1)(1|E)},
\end{equation}
which has the property that for any positive operator $E \in \Hb\otimes\Hb^*$
\begin{equation}
\bar{P}_j(E) \ge 0
\hspace{50pt}
\text{with equality if and only if}
\hspace{50pt}
E \propto \proj{j,\hat n}
\end{equation}
where the largest unirrep in $\Hb$ is equivalent to spin-$j$.
The spin-impurity is a homogeneous generalization of the invariant uncertainty for pure states in a spin-$j$ unirrep.
\begin{equation}
\Delta_j(\psi) \;=\; \langle \psi|\psi\rangle \langle \psi| \vec{J}^{\,2}| \psi\rangle - \langle \psi|\vec{J}\,|\psi\rangle^{2} \;=\; j^2\bar{P}\big(|\psi\rangle\!\langle \psi|\big) + j\!\braket{\psi}{\psi}^2  \;\ge\; j\!\braket{\psi}{\psi}^2.
\end{equation}

\section{The spin-coherent-state POVM via continuous isotropic measurement}\label{SCSPOVM}

This section is a review of the nonadaptive continuous isotropic measurement~\cite{shojaee2018optimal} followed by a better detailed proof of its effect as the spin-coherent-state POVM.
I begin with a formal discussion of continuous quantum measurement theory with emphasis on the ensemble interpretation of measurement outcome.
The nonadaptive continuous isotropic measurement is then introduced.
The unconditioned post-measurement state, $K K^\dag$, and effected POVM element, $K^\dag K$, are both calculated to be spin-coherent projectors with probability
\begin{equation}\label{guarantee}
\mathrm{Prob}\Big(\bar{P}\big(K^\dag K\big)< \epsilon\Big) = \mathrm{Prob}\Big(\bar{P}\big(KK^\dag\big) < \epsilon\Big) > 1 - \sqrt{\frac{6}{\pi\gamma T}}\,\ln\frac{2}{j\epsilon}
\end{equation}
where $q$ is the spin-impurity, equation \ref{spinimpurity}.

\subsection{General quantum measurement theory}

The building block will be a measurement modeled by coupling the system-of-interest with a Gaussian meter, via controlled displacement, which in turn is subjected to a homodyne measurement.
The choice of meter is to some extent arbitrary.
The predominant reason a Gaussian meter will be chosen is because the measurement of interest will be nonadaptive and continuous.
The central-limit-theorem tells us that whatever meter of a large class is fundamentally used, nonadaptivity will cause that meter in the continuum limit to behave effectively as a Gaussian meter.
Physical realizations which approximate the nonadaptive isotropic continuous measurement will have corrections that depend on the meter.

The Hilbert space for the system of interest will be symbolized by $\Hb_0$ with the state $\rho$ and observable $X$.
Measurements will be based on the standard model of coupling a meter to the system of interest and performing a von-Neumann measurement on the meter.
For meter, choose the unirrep of the Weyl-Heisenberg group,
\begin{equation}
[Q,P]=i,
\end{equation}
often referred to as a continuous-variable system.
Prepare the meter in a Gaussian pure state, with wavefunction in the position basis given by
\begin{equation}
\braket{q}{\psi} = \frac{1}{(2\pi \sigma^2)^{1/4}}e^{-q^2/2\sigma^2}.
\end{equation}
Interact the meter and system of interest with a bilinear Hamiltonian
\begin{equation}
	H t = \sqrt{\gamma t}\, \sigma P \otimes X.
\end{equation}
Subject the meter to a measurement of $Q$.
The Kraus operator is thus
\begin{equation}\label{standKraus}
	M(m)
	= \sqrt{\sigma\sqrt{\gamma t\,}\;}\Big\langle{q\!=\!m\sigma\sqrt{\gamma t}}\;\Big|e^{-iHt}\Big|\,\psi\Big\rangle
	= \left(\frac{\gamma t}{2\pi}\right)^{1/4}e^{-\frac{1}{4}\gamma t (X-m)^2}.
\end{equation}
The Hilbert-Schmidt inner product is of tremendous value.
Define
\begin{equation}
(A|B) = \Tr(A^\dag B)
\end{equation}
and
\begin{equation}
X\!\otimes\!Y^*|A) = |X\!AY^\dag).
\end{equation}

To report Kraus operators as in equation \ref{standKraus} is standard practice.
However, it is imperative to understand that this more accurately represents a \emph{quantum-operation-valued measure} (QOVM),
\begin{equation}
	d\Z(m) = dm\, M(m)\!\otimes\!M(m)^*.
\end{equation}
Interpret the associated trace-preserving superoperator
\begin{equation}
\Z = \int\!\!d\Z = e^{-\frac{1}{8}\gamma t (X\otimes 1 - 1 \otimes X^*)^2}
\end{equation}
as a partition function.
That $\Z$ is trace-preserving is symbolized by the identity
\begin{equation}
(1|\Z = (1|
\end{equation}
and similarly one recovers the usual POVM,
\begin{equation}
\big(dE(m)\big| = (1|d\Z = dm \left(M^\dag M\right|.
\hspace{30pt}
\end{equation}
To perform several measurements symbolized by partitions $d\Z_1$, $d\Z_2$, etc., the total QOVM is simply the composition
\begin{equation}
d\Z(\ldots, m_2, m_1) = \cdots d\Z_2(m_2) d\Z_1(m_1).
\end{equation}
This composition rule suggests a group-theoretic perspective.
Understand that the parameters in each $d \Z_k$ may be adaptive|that is, $d\Z_k$ may depend on the $d\Z_{l}(m_l)$ for $l<k$.

We are at liberty to separate a QOVM into an ``operator-valued part'' and a ``measure part'' in any way we conceptually see appropriate.
To parse the QOVM, a group-theoretic perspective insists one approach above others to be meaningful.
Define the classical Gaussian measure of zero mean,
\begin{equation}
d\mu(m) = \sqrt{\frac{\gamma t}{2\pi}}e^{-\frac{1}{2}\gamma tm^2},
\end{equation}
``linear Kraus operator'',
\begin{equation}
K(m) = e^{\frac{1}{2}\gamma t mX},
\end{equation}
and the operator
\begin{equation}
L(m) = e^{-\frac{1}{4}\gamma t X^2}K(m).
\end{equation}
The QOVM is thus
\begin{equation}
d \Z(m)
= d\mu(m)\,
L(m)\!\otimes\!L(m)^*
= e^{-\frac{1}{4}\gamma t X^2}\!\otimes\!e^{-\frac{1}{4}\gamma t  X^{*2}} d\mu(m)\,
K(m)\!\otimes\!K(m)^*
\end{equation}
and the POVM is
\begin{equation}
(dE| = (1|d\Z
= \left(L^\dag K\right| d\mu
= \left(e^{-\frac{1}{2}\gamma t X^2}K^\dag K\right| d\mu.
\end{equation}
The $K$ are particularly meaningful because if their $X$ represent a Lie algebra $\g$, then compositions or multiples of such $K$ are analytically closed|that is, for every $X,Y \in \g$, there is a $Z\in \g$ such that
\begin{equation}
	e^Ye^X = e^Z
\end{equation}
and such $K$ form a representation of the corresponding Lie group $G=e^\g$.
Similarly, understand that
\begin{equation}
K\!\otimes\!K^* = e^{\frac{1}{2}\gamma t m( X\otimes1+1\otimes X^*)}
\end{equation}
would also be a representation of $G$.
Multiplication of the $K$ is in general not physical, rather it is multiplication of the $L$ which represents the composition of measurements outputs.
Intervening nonlinearities are what we will call the factors of $e^{-\frac{1}{4}\gamma t X^2}$.

Taking the interaction strength to be infinitesimal, $ \gamma t \rightarrow \gamma dt \ll 1$,
Define the Wiener measure for a measurement record of duration time $T$,
\begin{equation}\label{Wmeas}
\D\mu[m] = \left(\frac{\gamma dt}{2\pi}\right)^{T/dt}e^{-\gamma\!\int_0^T\!dt \,m(t)^2},
\end{equation}
and define Wiener increments
\begin{equation}
dW(t) \equiv \gamma m(t) dt
\end{equation}
which by equation \ref{Wmeas} define an It\^o process, satisfying the It\^o rule
\begin{equation}
dW(t)dW(s) = \delta_{ts} \gamma dt.
\end{equation}
Consider a continuum of quantum operations
\begin{equation}
d\Z[m] = d\Z\big(m(dt), m(2dt),\ldots \big) = \cdots d\Z_2\big(m(dt)\big) d\Z_1\big(m(dt)\big) = \D\mu[m]\, L[m]\!\otimes\! L[m]^*.
\end{equation}
Let the measurement be nonadaptive, so then $L[m]=L(T)$ is the solution to the time-dependent stochastic differential equation
\begin{equation}\label{nonadaptequation}
dL(t) = \left(\frac{1}{2}X(t) dW(t) - \frac{1}{8}X(t)^2 \gamma dt \right)L(t)
\end{equation}
with initial condition $L(0)=1$.

In the stochastic setting, there is a subtlety between generators that are linear or nonlinear in the elements of a Lie algebra.~\cite{}
This is simply because
\begin{equation}\label{nointervening}
e^{\frac{1}{2}XdW} = 1 + X\frac{1}{2}dW + \frac{1}{8}X^2\gamma dt.
\end{equation}
The systematic term of equation \ref{nointervening} must be understood as required to keep the nondifferentiable increment, $XdW$, in the group generated by $X$.
One must take extra care to make sure the systematic term of equation \ref{nointervening} is not confused with an intervening nonlinearity such as in equation \ref{nonadaptequation}.

\subsection{The isotropic continuous measurement}\label{isomeasure}

Having our apparatus of techniques, I now review the isotropic continuous measurement.
The protocol is very simple, just measure $J_x$, $J_y$, and $J_z$ continuously and with equal strength.
The QOVM is thus generated by the weak QOVM
\begin{equation}
	d\Z(\vec m) = d\Z(m^z)d\Z(m^y)d\Z(m^x) = d\mu(\vec{m})\, L_{3dt}(\vec m)\!\otimes\! L_{3dt}(\vec m)^*
\end{equation}
where
\begin{equation}
d\mu(\vec{m}) = \left(\frac{\gamma dt}{2\pi}\right)^{3/2}e^{-\frac{1}{2}\gamma dt{\vec m}^2}
\end{equation}
and most importantly
\begin{equation*}
L_{3dt}(\vec m) =
e^{\frac{1}{2}J_zdW^z-\frac{1}{4}J_z^2\gamma dt}
e^{\frac{1}{2}J_ydW^y-\frac{1}{4}J_y^2\gamma dt}
e^{\frac{1}{2}J_xdW^x-\frac{1}{4}J_x^2\gamma dt}
\end{equation*}
\begin{equation}
= e^{\frac{1}{2}\vec{J}\cdot d\vec{W}-\frac{1}{4}\vec{J}^{\,2}\gamma dt}.
\end{equation}
The crucial observation is that the quadratic generator becomes proportional to a Casimir operator and thus no longer intervenes with any linear generator because
\begin{equation}
[\vec{J}^{\,2},J_k]=0.
\end{equation}
Continuing this isotropic measurement for a time $T$, the QOVM is thus a path integral of the measurement record
\begin{equation}\label{contmeas}
\D\Z[\vec m]
= d\Z\big(\vec{m}(T)\big)\cdots\,d\Z\big(\vec{m}(6dt)\big)d\Z\big(\vec{m}(3dt)\big)
= e^{-\frac{\gamma T}{12}\vec{J}^{\,2}} \!\!\otimes\! e^{-\frac{\gamma T}{12}\vec{J}^{\,2}} \D\mu[\vec{m}]\; K[\vec{m}] \!\otimes\! K[\vec{m}]
\end{equation}
where $\K[\vec{m}]$ is the solution to the stochastic differential equation
\begin{equation}\label{finalevolution}
dK(t) = \left(\frac{1}{2}\vec{J}\!\cdot\!d\vec{W} + \frac{1}{8}\vec{J}^{\,2}\gamma dt\right)K(t).
\end{equation}

\subsection{Effect of the nonadaptive, isotropic, continuous measurement}

Define \emph{spin-coherent operators} (SCOs) for a spin-$j$ unirrep as operators of the form
\begin{equation}
U(R)\proj{j}U(S)^\dag = \ketbra{j,\hat n}{j, \hat m}e^{-ij(\psi_R-\psi_S)} .
\end{equation}
What remains to be shown is that for the Wiener ensemble, equation \ref{Wmeas}), every measurement record over a long enough time is guaranteed to give an SCO for a Kraus operator in the sense of equation \ref{guarantee}.
Indeed, this is a direct property of the stochastic differential equation \ref{finalevolution}.

By closure of the group, we know that the solution $K(t)$ to equation \ref{finalevolution} at every time is a representation of the complexification of $\SU(2)$, which is isomorphic to $\SL_\C(2)$.
In particular, this means that the singular-value decomposition of $K$ is such that
\begin{equation}
K(t) = U(t)e^{A(t)}V(t)^\dag
\end{equation}
where $U$ and $V$ are representations of $\SU(2)$ and $A = \alpha J_z$ is Hermitian diagonal.
Since, $U$ and $V$ are representations of $\SU(2)$ there exist stochastic $i\,\dbar \psi$ and $i\,\dbar \phi$ and ballistic $i\zeta dt$ and $i\omega dt$ which are representations of $\su(2)$ such that
\begin{equation}
	U^\inv d\,U = i\,\dbar{\psi} - \frac{1}{2}\,\dbar{\psi}^2+ i \xi dt
\hspace{50pt}
\text{and}
\hspace{50pt}
	V^\inv dV = i\,\dbar{\phi} - \frac{1}{2}\,\dbar{\phi}^2  + i\omega dt.
\end{equation}
A straightforward calculation shows
\begin{equation*}
	dK K^\inv =
	 U \bigg[ U^\inv d\,U
	 +\left(dA + \frac{1}{2}dA^2\right)
	- e^A\Big(V^\inv dV\Big)e^{-A}
	\hspace{100pt}
\end{equation*}
\begin{equation}
\hspace{100pt}
	+ U^\inv d\,U dA
	- dA e^{A}\Big(V^\inv dV\Big)e^{-A}
	-U^\inv d\,U e^{A}\Big(V^\inv dV\Big)e^{-A} \bigg]U^\inv
\end{equation}
\begin{equation*}
= U \bigg[ i\Big(\,\dbar \psi + \xi dt - \cosh\ad_A\big(\,\dbar\phi+\omega dt\big)\Big)
+\, dA  - i\sinh\ad_A\big(\,\dbar\phi+\omega dt\big)
\end{equation*}
\begin{equation}\label{compareto}
\hspace{150pt}
+\,\frac{1}{2}\Big(dA +i\,\dbar\psi - i\cosh\ad_A\big(\,\dbar\phi) - i\sinh\ad_A\big(\,\dbar\phi)\Big)^2
\bigg]U^\inv.
\end{equation}
Define $\pi$ to be the projector onto the one-dimensional algebra spanned by $J_z$ and $\bar{\pi}$ to be the projector onto the complement spanned by $J_x$ and $J_y$.
Comparing equation \ref{compareto} with equation \ref{finalevolution} and equating linearly independent terms gives
\begin{equation}
dA  = \frac{1}{2}\pi\Big(U^\inv \!J_\mu U\Big)dW^\mu,
\end{equation}
\begin{equation}
-i\sinh\ad_A\big(\,\dbar\phi\big) = \frac{1}{2}\bar{\pi}\Big(U^\inv \!J_\mu U\Big)dW^\mu,
\end{equation}
\begin{equation}
\dbar \psi = \cosh\ad_A\big(\,\dbar\phi\big),
\end{equation}
and
\begin{equation}\label{gauge}
\sinh\ad_A\big(\omega\big) = 0
\hspace{50pt}
\text{and}
\hspace{50pt}
\xi = \cosh\ad_A\big(\omega\big).
\end{equation}
Equations \ref{gauge} tell us that
\begin{equation}
\bar{\pi}(\omega) = \bar\pi(\xi)= 0
\hspace{50pt}
\text{while}
\hspace{50pt}
\pi(\omega) = \pi(\xi)
\end{equation}
is a gauge degree of freedom corresponding to the symmetry
\begin{equation}
	(U,V) \longrightarrow 	(Ue^{-i\theta J_z},Ve^{-i\theta J_z})
\end{equation}
which we set equal to zero.
Similarly, $\pi(\,\dbar \psi) = \pi(\,\dbar \phi)$ is a gauge degree of freedom we set equal to zero.
In summary, we have
\begin{equation}\label{final0}
dA = \frac{1}{2}\pi\Big(U^\inv \!J_\mu U\Big)dW^\mu,
\end{equation}
\begin{equation}\label{final1}
-i\,\dbar\phi = \frac{1}{2}\csch\,\ad_A\bigg(\bar{\pi}\Big(U^\inv \!J_\mu U\Big)\!\bigg)dW^\mu,
\end{equation}
and
\begin{equation}\label{final2}
-i\,\dbar \psi = \frac{1}{2}\coth\ad_A\bigg(\bar{\pi}\Big(U^\inv \!J_\mu U\Big)\!\bigg)dW^\mu.
\end{equation}

Several important properties of $K(t)$ can be derived from equations \ref{final0} through \ref{final2}.
First of all, one should observe that $A = \alpha J_z$ is a Gaussian random variable with zero mean and variance
\begin{equation}
	\Big\langle \alpha(T)^2\Big\rangle = \frac{\gamma T}{12}
\end{equation}
where angular brackets denote the ensemble average over measurement records.
In particular, this means that $\alpha$ is guaranteed to increase without bound in the sense that for any finite $\alpha_o$
\begin{equation}\label{prob}
\mathrm{Prob}\Big(|\alpha(T)|\!<\!\alpha_o\Big) < 2\alpha_o\sqrt{\frac{6}{\pi\gamma T}}.
\end{equation}
In turn, this means that by equation \ref{final1} the POVM element
\begin{equation}\label{guarantee1}
\D\mu[m] E(T) = e^{-\frac{\gamma T}{6}\vec{J}^{\,2}}\D\mu[m] V(T)e^{2\alpha(T)J_z}V(T)^\dag
\end{equation}
is only a function of the beginning of the measurement record while the unconditioned post-measurement state proportional to
\begin{equation}\label{guarantee2}
K(T)K(T)^\dag = U(T)e^{2\alpha(T)J_z}U(T)^\dag
\end{equation}
continues to evolve indefinitely.
In fact, one should observe that by equation \ref{contmeas} could just as well think of $K[\vec{m}]$ as the solution to a similar differential equation as \ref{finalevolution} except integrated backwards in time with $K(t)$ on the left.  From this observation one concludes that $U(T)$ is only a function of the end of the measurement record.

Finally, make the observation that equations \ref{guarantee1} and \ref{guarantee2} are guaranteed to become projectors onto coherent states.
A simple calculation reveals
\begin{equation}
\bar{P}_j(K^\dag \!K) = \bar{P}_j(K \!K^\dag )
= 1 - \left(\frac{\big(J_z\big|e^{2\alpha J_z}\big)}{j\big(1\big|e^{2\alpha J_z}\big)}\right)^2 < \frac{2}{j}e^{-2|\alpha|}.
\end{equation}
Combining this with inequality \ref{prob} gives the bounds at the beginning of this section, namely inequality \ref{guarantee}.

\section{Generalized-coherent-state POVMs via continuous isotropic measurement}\label{GCSPOVM}

This section introduces generalized-coherent states, the generalized-coherent-state POVM, the continuous isotropic measurement, and a proof that the continuous isotropic measurement implements the generalized-coherent-state POVM.
To develop the required perspective and techniques, I begin with a description of the geometry of complex semisimple Lie groups relative to their maximal compact subgroups.
This is precisely the structure of Kraus operators encountered in the measurement of any finite-dimensional unitary representation of a Lie group.
This relative geometry is most simply a perspective on the singular-value decomposition as a coordinate system for the group of Kraus operators.

Although the components of this discussion for the generalized-coherent-state POVM each parallel the discussion for the spin-coherent-state POVM, I will cover these components in a rather different order.
After describing semisimple Lie groups and their representations, I will immediately discuss the properties of the general stochastic differential equations analogous to equation \ref{finalevolution}.
This is in part to emphasize the representation independent aspects of the result.
The section then ends with a quick description of the continuous isotropic measurement, generalizing slightly the details by which one may achieve isotropy.

\subsection{Compact semisimple Lie groups and their representations}\label{CSS}

The purpose of this section is to describe of the spectrum for a maximal commuting set of observables representing the generators of a compact Lie group.
Such observables can be simultaneously diagonalized by the very same group they generate.
Their simultaneous eigenvalues, called \emph{weights}, are quantized with spacings described by generalized ladder operators.
The state of highest weight is nondegenerate and uniquely defines the unirrep.
Unlike for noncompact Lie groups, every raising operator of a compact Lie group can be transformed into its conjugate lowering operator by group conjugation.
This results in each spectrum having plenty of reflection symmetries, known as Weyl reflections.

\vspace{15pt}

Every compact connected Lie group is the Cartesian product of a semisimple Lie group and finitely many commuting phases.
Let us ignore these phases.
Semisimple Lie groups are groups which have Lie algebras that can be decomposed into a maximal commuting set of generators called a Cartan subalgebra (CSA) and (with complex coefficients) conjugate pairs of raising/lowering operators, called the Cartan-Weyl basis.
Compact semisimple Lie groups have only finite-dimensional unirreps and every such unirrep has a basis of CSA eigenstates, of which the simultaneous eigenvalues are called weights.
The state of highest weight is the unique, non-degenerate state which is annihilated by all raising operators.

\begin{center}
	\begin{framed}[0.8\textwidth]
		Let $G_o$ be a compact connected semisimple Lie group and let $\g_o$ be its Lie algebra.
	\end{framed}
\end{center}

The choice of CSA is arbitrary as well as is the choice of which in each conjugate pair of ladder operators is the raising operator.
Nevertheless, each CSA of a $\g_o$ is equivalent to every other CSA under conjugation by an element of $G_o$.
In particular, this means every element of $\g_o$ is conjugate (by an element of $G_o$) to an element of a fixed CSA.
This is a generalization of the familiar spectral theorem that applies when $G_o\cong\SU(d)$: the fixed CSA is the generalization of diagonal matrices and group conjugation is the generalization of a similarity transform.
This orbit (under representations of $G_o$) of the highest-weight state is the set of $G_o$-coherent states (GCSs.)
In particular, this means that the ground state of any regular (defined after equation \ref{reg}) Hamiltonian representing an element of $\g_o$ is a nondegenerate GCS.
Hamiltonians representing an element of $\g_o$ we may call linear as every Hamiltonian over the Hilbert space is a polynomial in the elements of $\g_o$.

\begin{center}
	\begin{framed}[0.75\textwidth]
		Let $\h_o$ be a choice of CSA in $\g_o$ and $\Delta_+$ be the choice of raising operators so\\ $\g_o = \h_o \oplus \bigoplus_{\alpha \in \Delta_+} (\g_\alpha \oplus \g_{-\alpha})$.
	\end{framed}
\end{center}

The dimension $r = \dim\h_o$ is called the \emph{rank} of $G_o$.
Suppose $\{H_k\}$ is a basis of $\h_o$.
Each $\g_\alpha$($\g_{-\alpha}$) is a linear space spanned by a single raising (lowering) operator $L_\alpha$($L_{-\alpha}$); these operators are defined by their simultaneous eigenvalues $\{\alpha_k\}$ under the adjoint representation of  $\{H_k\}$,
\begin{equation}
\ad_{H_k}L_{\pm\alpha} \equiv [H_k,L_{\pm\alpha}] = {\pm\alpha}_k L_{\pm\alpha}.
\end{equation}
One can think of each $\alpha$ as representing an $r$-tuple $\alpha = \{\alpha_k\}$.
However, it is far more convenient to think of each $\alpha$ as a linear function $\alpha:\h_o \longrightarrow \R$ since for any $H = \gamma^k H_k \in \h_o$ we have
\begin{equation}
\ad_{H}L_{\pm\alpha} = \ad_{\gamma^k H_k}L_{\pm\alpha} = ({\pm\alpha}_k\gamma^k) L_{\pm\alpha}= \pm\alpha(H) L_{\pm\alpha}.
\end{equation}
These linear functionals are called the \emph{roots} of $\g_o$ and this name is because for any $X= gHg^\inv\in\g_0$,
\begin{equation}\label{reg}
\det\big(z-\ad_X\big) = \det\big(z-\ad_H\big) = z^r \prod_{\alpha \in \Delta_+}\big(z-\alpha(H)\big)\big(z+\alpha(H)\big).
\end{equation}
Regular elements of $\g_o$ are elements such that $\alpha(H)\neq 0$ for any $\alpha$.

The ladder operators come in pairs and the choice of which is the raising operator is better referred to as the choice of positive roots $\alpha \in \Delta_+$.
Just as the structure of the CSA choice is decribed by their equivalence via group conjugation, the choice of positive roots is also described by reflection symmetries via group conjugation:
For each positive root $\alpha$ there is an element $w_\alpha \in G_o$ such that
\begin{equation}
w_\alpha H w_\alpha^\inv = H - 2\frac{\alpha(H)}{(\alpha,\alpha)}H_\alpha
\end{equation}
where
\begin{equation}
H_\alpha = g^{jk}\alpha_jH_k
\hspace{10pt}
\text{,}
\hspace{90pt}
(\alpha,\beta) = g^{jk}\alpha_j\beta_k
\hspace{10pt}
\text{,}
\end{equation}
and $g^{jk}$ is the inverse of the restricted Killing form
\begin{equation}
g_{jk} = \Tr\,\ad_{H_j}\ad_{H_k}.
\end{equation}
In particular, this means
\begin{equation}
w_\alpha L_\alpha w_\alpha^\inv = L_{-\alpha}
\hspace{50pt}
\text{and}
\hspace{50pt}
w_\alpha L_\beta w_\alpha^\inv = L_{W_\alpha(\beta)}
\end{equation}
where
\begin{equation}
W_\alpha(\beta) = \beta - 2\frac{(\alpha,\beta)}{(\alpha,\alpha)}\alpha.
\end{equation}
The $\{W_\alpha\}$ define the abstract Weyl group of the Lie algebra $\g_o$ and choices of $\{w_\alpha\}$ are known as analytic representations of the Weyl group.
In general, analytic representations of the Weyl group are projective with fiber isomorphic to the maximal torus $T_o \equiv e^{\h_o} \subset G_o$.
These reflections divide the CSA up into conjugation equivalent regions known as Weyl chambers.

\begin{center}
	\begin{framed}[0.7\textwidth]
		Let $\Hb$ be a Hilbert space that carries the unirrep $U:G_o\longrightarrow\SU(\Hb)$\\
		and the associated irreducible representation $-iJ:\g_o\longrightarrow\su(\Hb)$.
	\end{framed}
\end{center}

Such Hilbert spaces $\Hb$ afford a basis of simultaneous eigenstates of $J(\h_o)$,
\begin{equation}
J(H)\ket{\mu} = \mu(H) \ket{\mu}.
\end{equation}
These simultaneous eigenvalues, represented as linear functionals are called the weights of $\Hb$, $U$, or $J$.
Let us define
\begin{equation}
J_{\pm\alpha} = J(L_{\pm\alpha})
\hspace{50pt}
\text{and}
\hspace{50pt}
U_\alpha = U(w_\alpha).
\end{equation}
Under these ladder operators and Weyl reflections, we have
\begin{equation}
J_{\pm\alpha}\ket{\mu} \propto \ket{\mu\pm\alpha}
\hspace{50pt}
\text{and}
\hspace{50pt}
U_\alpha \ket{\mu} \propto \ket{W_\alpha(\mu)}.
\end{equation}
If $\mu$ and $\nu$ are two weights, then $\nu-\mu = n\alpha$ for some integer $n$ and root $\alpha$.
Further, $\mu + k\alpha$ is a weight for every integer $k$ between $0$ and $n$.
Finally, there is a unique, non-degenerate state of highest weight such that
\begin{equation}
J_\alpha\ket{\lambda}=0
\end{equation}
for every positive root $\alpha \in \Delta_+$.
Finally, a $G$-coherent state (GCS) is any state in the orbit
\begin{equation}
G^\lambda = \big\{U(g)\ket{\lambda} : g \in G_o\big\}.
\end{equation}
One can also define $G_o$-coherent Gibbs states, $\rho = e^{-\beta F}$ for any linear $F \in J(\g_o)$.
Remember that every Hamiltonian in $\su(\Hb)$ is a polynomial in the elements of $J(\g_o)$ and that every regular linear Hamiltonian has a non-degenerate GCS for its groundstate.
The GCS POVM is defined by
\begin{equation}
dE(g) = d\mu(g) U(g)\proj{\lambda} U(g)^\dag
\end{equation}
where $d\mu(g)$ is the Haar measure of $G_o$, properly normalized so that $\int\!\!dE(g) = \boldsymbol{1}$.

\begin{figure}[h!]
	\centering
	\includegraphics[height=2.0in]{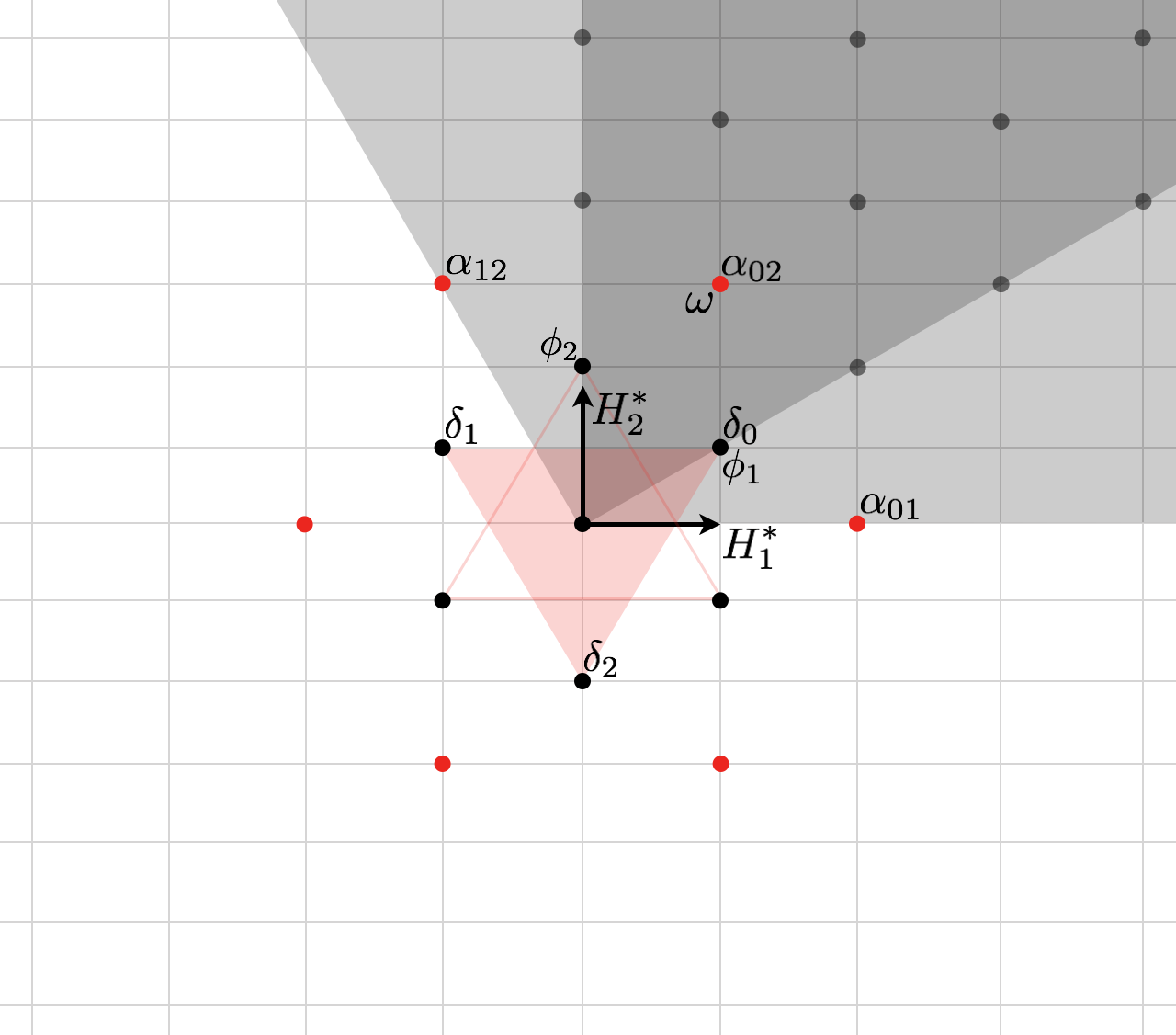}
	\hspace{50pt}
	\includegraphics[height=2.0in]{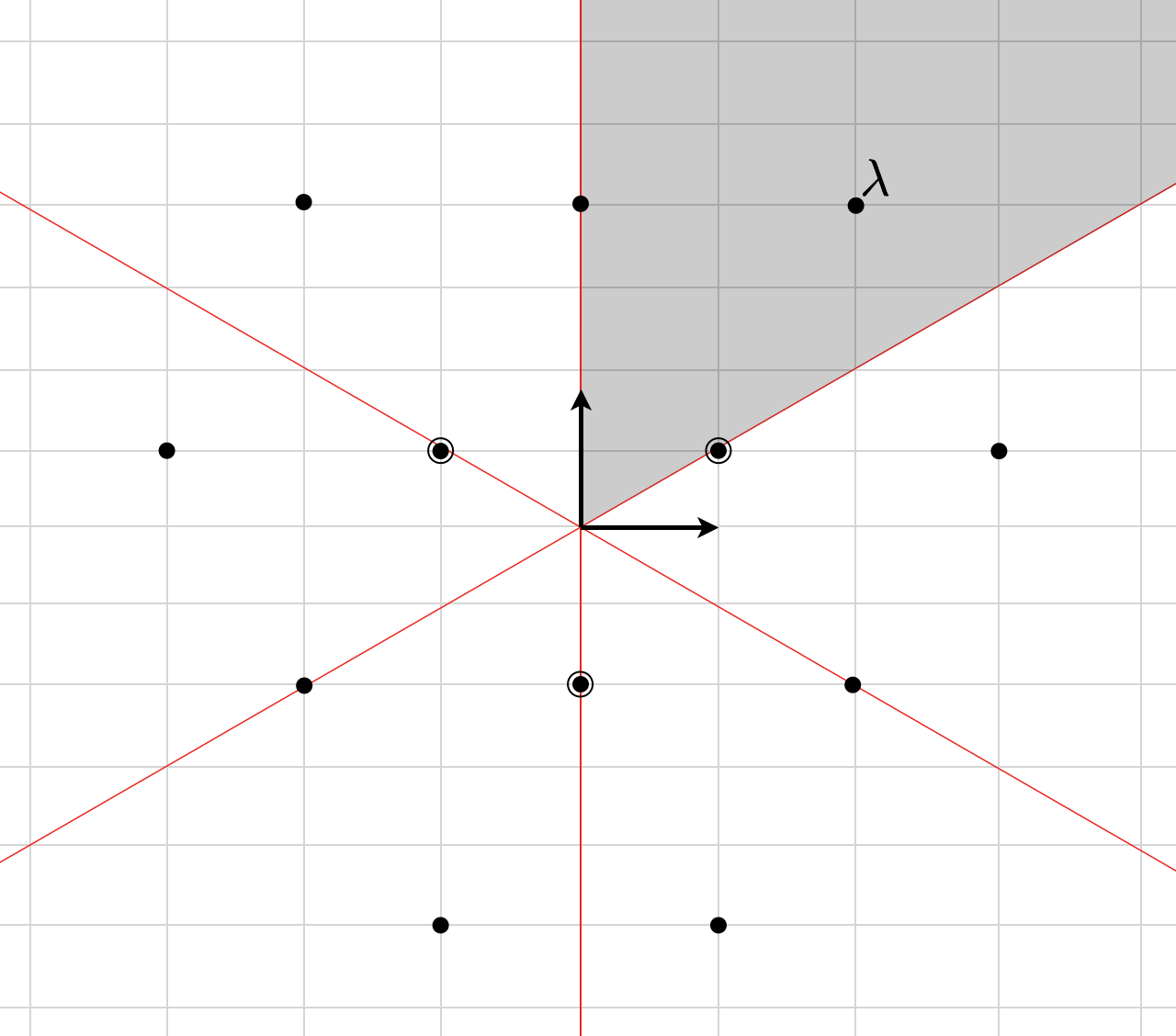}
	\caption{
		Weights/representations of $\SU(3)$:
		Let $\{\ket0, \ket1, \ket2\}$ be an orthonormal basis of the vector space carrying the defining representation.
		A Cartan subalgebra $\h$ is the algebra of diagonal traceless matrices of which $\{H_1=\proj{0}-\proj{1},H_2=(\proj{0}+\proj{1}-2\proj{2})/\sqrt{3}\}$ is a basis orthonormal under $(X|Y)/2$.
		The black arrows, $\{H_1^*,H_2^*\}$, are the basis of $\h^*$ dual to this orthogonal basis of $\h$.
		The red dots are the roots corresponding to the Weyl generators $\{E_{\alpha_{kl}}=\ketbra{k}{l}:k \neq l\}$.
		$\{\alpha_{01},\alpha_{02},\alpha_{12}\}$ are a choice of positive roots with primitive roots $\{\alpha_{01},\alpha_{12}\}$ and positive cone the wider gray region.
		The narrow gray region is the Weyl chamber and $\{\phi_1,\phi_2\}$ are the fundamental weights.
		The weights of the defining representation are $H\ket{k}=\delta(H)\ket{k}$.
		The red lines are the planes of reflection corresponding to elements of the Weyl group.
		On the right is an irrep with the 15-dimensional irrep of highest weight $\lambda = 2\phi_1+\phi_2$.
		The nine weights on the border have no degeneracy, while the three weights in the interior each have two-fold degeneracy.}
	\label{highestweights}
\end{figure}

\subsection{Effect of the continuous isotropic measurement}\label{SVD}

Up until now, we have been somewhat relaxed about observables, Hermitian operators, generators of $G_0$, and ladder operators.
However, these distinctions turn out to be very important in the context of modelling measurement and so we must be more careful.
It is standard in quantum physics to refer to a Hermitian operator $J$ as the generator of a unitary $U=e^{-i\theta J}$.
However, in this section I will refer to the generator of the unitary $U=e^{-i\theta J}$ as the anti-Hermitian operator, $-iJ$.
To be clear, this convention will only be a linguistic distinction as all anti-Hermitian operators will still be denoted as $-iJ$ where $J$ is Hermitian.
Kraus operators will be generated by the complex semisimple Lie algebra
\begin{equation}
\g = \g_0 \oplus i\g_0
\end{equation}
and thus be representations of the complex semisimple Lie group
\begin{equation}
G = e^{\g}.
\end{equation}
As observables, Hermitian operators will be associated with measurements and considered elements of the subspace $i\g_0$.
Displacements generated by $i\g_0$ will be stochastic, corresponding to the natural randomness in the outcomes of quantum measurement.

A very useful fact is that $G$ has a generalized singular-value decomposition\footnote{This is referred to in the literature as the ``$KAK$-theorem'' for type-IV globally Riemannian symmetric spaces.} relative to $G_0$.
This decomposition is such that every $k \in G$ is of the form
\begin{equation}\label{svd}
k = ue^{a}v^\dag
\end{equation}
where $u,v \in G_o$ are the frames and $a \in i\h_o$ is the generalized singular value.
This decomposition will prove useful as a coordinate system as we will want to parse the evolution of stochastic differential equations (SDEs) of the form
\begin{equation}\label{gito}
dk(t) = \left(\frac{1}{2}\,\dbar w(t)+\frac{1}{8}\,\dbar w(t)^2\right) \,k(t)
\end{equation}
where $\dbar w$ is an $i\g_o$-valued Wiener increment, isotropic under the Killing form
\begin{equation}
B(X,Y) = \Re\big(\Tr\, \ad_X \ad_Y\big).
\end{equation}
What makes equation \ref{svd} so useful is that under the Killing form, $\g_o$ and $i\g_o$ are orthogonal and $\h_o$ is orthogonal to the ladder operators.
Indeed, a quick calculation reveals
\begin{equation}\label{neat}
dk k^\inv = u\Big(u^\inv du - \cosh{\ad_a}(v^\inv dv)\Big)u^\inv + u\Big(da - \sinh{\ad_a}(v^\inv dv)\Big)u^\inv + \text{systematic terms}
\end{equation}
where the first term is in $\g_o$, the second in $i\g_o$, $da \in i\h_o$, and $\sinh{\ad_a}(v^\inv dv)$ is orthogonal to $i\h_0$.
Let $\pi$ be the projection onto the CSA and $\overline{\pi} = 1 - \pi$ be the orthogonal projection.
Equating \ref{neat} with the SDE \ref{gito} gives for the evolution of the singular value
\begin{equation}\label{first}
da = \frac{1}{2}\pi\big(u^\inv \dbar w u\big)
\end{equation}
for the evolution of the right frame
\begin{equation}\label{second}
v^\inv dv = -\frac{1}{2}\csch\,\ad_a\overline{\pi}\big(u^\inv \dbar w u\big)
\end{equation}
and for the evolution of the left frame
\begin{equation}\label{third}
u^\inv du = \cosh{\ad_a}\big(v^\inv dv\big) = -\frac{1}{2}\coth\,\ad_a\overline{\pi}\big(u^\inv \dbar w u\big).
\end{equation}

Let $\{X_\mu\}$ be a Hermitian basis of $\g$, define
\begin{equation}
g_{\mu\nu} = B(X_\mu,X_\nu),
\end{equation}
and let $\dbar w = X_\mu dW^\mu$ so that
\begin{equation}
dW^\mu dW^\nu = g^{\mu\nu}\gamma dt.
\end{equation}
It is easy to see that equation \ref{first} implies that $a(T) = \alpha^i(T)H_i$ will increase without bound as a zero-mean Gaussian random variable with variance
\begin{equation}\label{genprob}
\big\langle \alpha^k(T)\alpha^l(T) \big\rangle = \frac{\gamma Tg^{kl}}{4\dim{\g_o}}.
\end{equation}
This diffusive increase further causes the right frame to stop evolving by equation \ref{second} and the left frame to wander forever by equation \ref{third}.
By the symmetry of equation \ref{gito},
\begin{equation}\label{bito}
dk = k \Ad_k(\,\dbar w),
\end{equation}
we can imagine integrating backwards in time to also see that the left frame is only a function of the latest part of the Wiener process.

Physically, this analysis should makes complete sense since the Wiener process represents a measurement record and the maps
\begin{equation}
k \longmapsto k^\dag k = v e^{2f} v^\dag
\hspace{50pt}
\text{and}
\hspace{50pt}
k \longmapsto k k^\dag = u e^{2f} u^\dag
\end{equation}
represent the POVM element and unconditioned post-measurement state, respectively.
Indeed, this analysis says that the unconditioned post-measurement state will only depend on the latest part of the measurement record and thus be uncorrelated with the initial state due to projection noise.
On the other hand, the measurement outcome will only be a function of the earliest part of the measurement record representing the usual collapse due to back-action.
In the context of a representation of $G$, this increase in the singular value has yet another effect.

\begin{center}
	\begin{framed}[0.9\textwidth]
		Let $G$ be the complex semisimple Lie group with maximal compact subgroup $G_o$ and\\ $K:G\longrightarrow\SL_\C(\Hb)$ be the irreducible representation extended from the unirrep $U:G_o\longrightarrow\SU(\Hb)$.
	\end{framed}
\end{center}
Define \emph{$G$-coherent operators} (GCOs) to be operators in
\begin{equation}
G^\lambda \otimes G^{\lambda^*} = \big\{ z U(g)\proj{\lambda}V(h)^\dag : g,h \in G_0 \;\text{ \& }\; z\in\C \big\}.
\end{equation}
Every measurement record sampled from a Wiener process is guaranteed to be such that the solution to equation \ref{gito} limits to a GCO.
This can be demonstrated with a generalized $G$-impurity
\begin{equation}
\bar{P}_\lambda(E) = 1- \frac{(E|J_\mu)g^{\mu\nu}(J_\nu|E)}{(E|1)\,|\lambda|^2(1|E)}
\end{equation}    
where $|\lambda|^2 = (\lambda,\lambda)$.
This $G$-impurity also has the property that for any positive operator $E \in \Hb\otimes\Hb^*$
\begin{equation}
\bar{P}_\lambda(E) \ge 0
\hspace{50pt}
\text{with equality if and only if}
\hspace{50pt}
E \propto U(g)\proj{\lambda} U(g)^\dag.
\end{equation}
Indeed, letting $A = J(a) = \beta^iJ(H_i)$, an application of the Weyl character formula reveals
\begin{equation}\label{genimp}
\bar{P}_\lambda(K^\dag K) = \bar{P}_\lambda(KK^\dag)
= 1 - \frac{\big(e^{2A}|J(H_k)\big)g^{kl}\big(J(H_l)|e^{2A}\big)}{\big(e^{2A}|1\big)|\lambda|^2\big(1|e^{2A}\big)}
< \frac{4}{|\lambda|^2}\big(\lambda,\omega\big)e^{-4\omega(|a|)}
\end{equation}
where $|a|$ is the reflection of $a$ contained in the positive Weyl chamber.
Once again, equations \ref{genprob} and \ref{genimp} together in turn give us the guarantee that
\begin{equation}\label{generalguarantee}
\mathrm{Prob}\Big(\bar{P}\big(K^\dag K\big)< \epsilon\Big) = \mathrm{Prob}\Big(\bar{P}\big(KK^\dag\big) < \epsilon\Big) > 1 - \sqrt{\frac{2\dim\g_o}{\pi\gamma T}}\,\ln\frac{4(\omega,\lambda)}{|\lambda|^2\epsilon}.
\end{equation}

\begin{figure}[h!]
	\centering
	\includegraphics[height=2.5in]{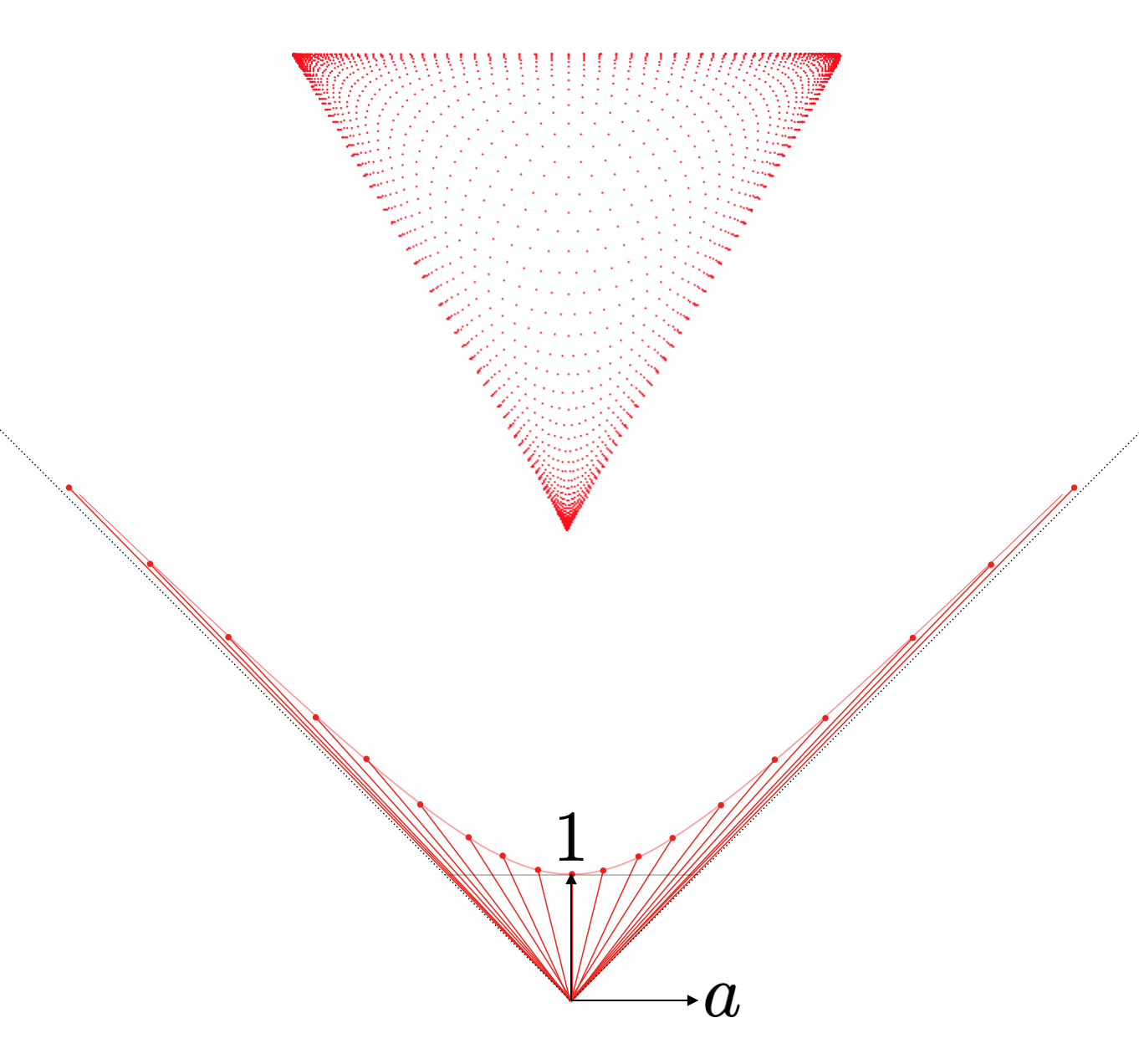}
	\hspace{50pt}
	\includegraphics[height=2.5in]{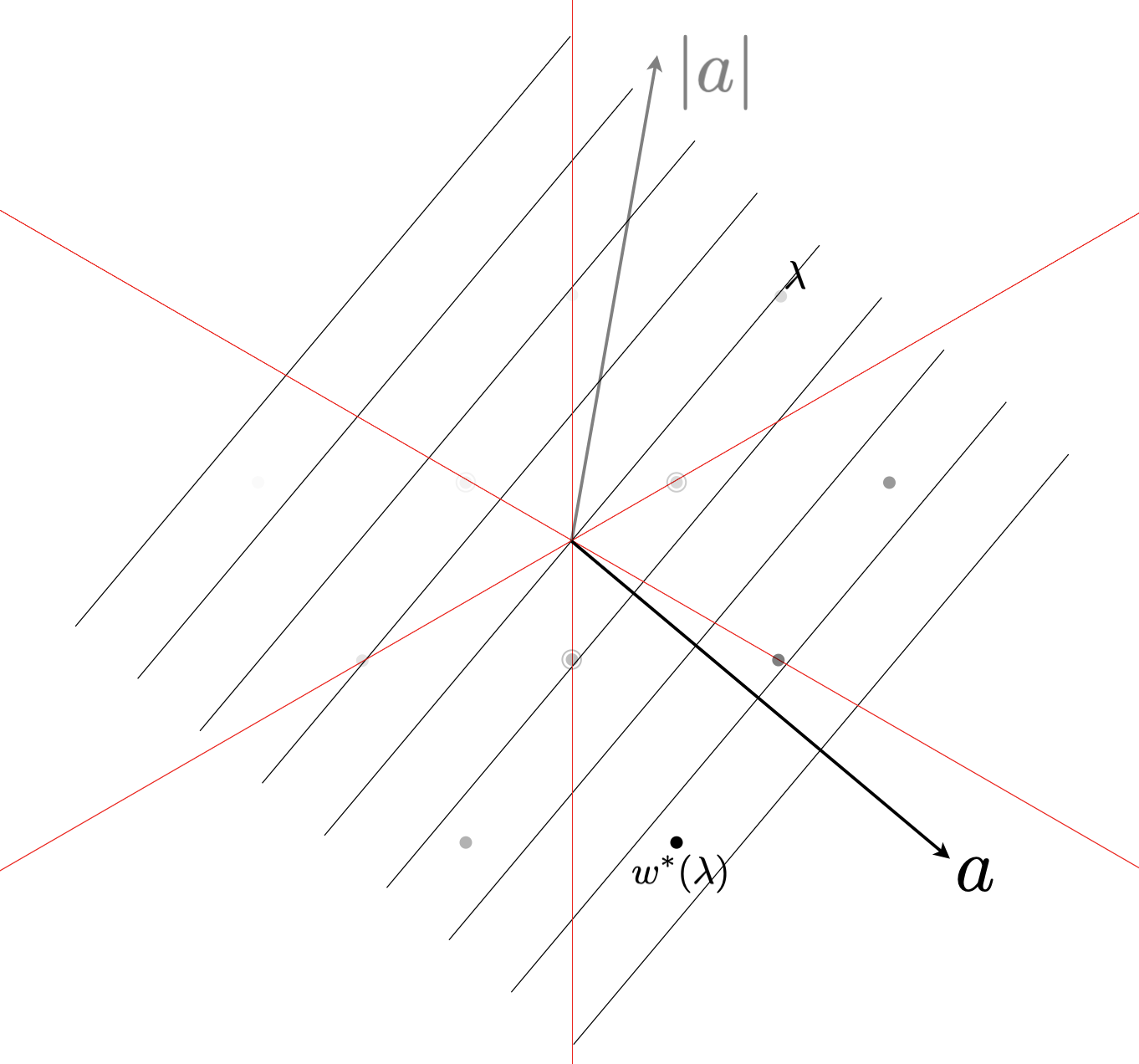}
	\caption{
		On the bottom left is the hyperbola $e^{2a} = e^{\alpha \sigma_z}$ of $\SU(2)$ singular values.  Every representation of this hyperbola is diffeomorphic to the hyperbola.  The rays are meant to help visualize the fact that a uniform distribution on the hyperbola translates to a singular distribution of rays as shown by their intersection with a normalized plane.
		On the top left is the normalized plane for the defining representation of $\SU(3)$.
		The red dots are the intersection of the plane with rays pointing to a uniform set of points on the parametric surface $e^{2a}$ for $\SU(3)$.
		On the right is a randomly sampled $a$ over the representation given by the same weight diagram as in the 15-dimensional example of figure \ref{highestweights}.
		The opacity of each weight $\mu$ is a visualization of the magnitude of the coefficients $e^{\mu(a)}$.}
	\label{weights}
\end{figure}

\subsection{The continuous isotropic  measurement}\label{complexgroups}

Let the system-of-interest be a Hilbert space $\Hb_o$ which carries a unirrep of the compact Lie group $G_o = e^{\g_o}$.
In the continuum limit, if the representation is subjected to finitely many measurements, then over such an infinitesimal time there will not yet be any non-abelian effects.
Let $n$ observables $\{X_a\}$ be coupled to $p$ pointer states with momenta $\{P_\alpha\}$, and $\{(g\tau)^{a\alpha}\}$ be their coupling strengths and durations.
Perpare the meter in a joint Gaussian state
\begin{equation}
\langle q^\alpha|\boldsymbol\sigma^2\rangle
= \exp\left(-\frac{1}{4}(\boldsymbol\sigma^{-2})_{\alpha\beta}q^\alpha q^\beta\right)
\end{equation}
and interact this meter with the representation by a Hamiltonian
\begin{equation}
Hndt = \sum_{a,\alpha} (g\tau)^{a\alpha} X_a\otimes P_\alpha
\end{equation}
then the weak Kraus operator effected on the representation would be
\begin{equation}
M(\{q^\alpha\})  = \exp\left(-\frac{1}{4}(\boldsymbol\sigma^{-2})_{\alpha\beta}\Big(q^\alpha-(g\tau X)^\alpha\Big) \Big(q^\beta-(g\tau X)^\beta\Big)\right).
\end{equation}

Let $K : G \rightarrow \SL_\C(\Hb_o)$ be the associated representation of the complex semisimple group $G = e^\g = e^{\g_0 \oplus i\g_0}$.
Let $\{-iX_\mu\}$ be a basis for $K(\g_0)$ and ${c_a}^\mu$ be the coefficients of the measured observables
\begin{equation}
X_a = {c_a}^\mu X_\mu.
\end{equation}
Gather the parameters for the interaction between the pointer observables $\{P_a\}$ and the basis of primary observables $\{X_\mu\}$ into a total effective coupling tensor
\begin{equation}
\kappa^{\alpha\mu}= (g\tau)^{a\alpha}{c_a}^\mu.
\end{equation}
For $n,p\ge\dim\g$, one can design these parameters to be isotropic, such that they satisfy the condition
\begin{equation}
(\boldsymbol\sigma^{-2})_{\alpha\beta}\kappa^{\alpha\mu}\kappa^{\beta\nu} = n \gamma dt \frac{g^{\mu\nu}}{\dim \g_o}.
\end{equation}
It is important to notice that the compactness of the Lie group plays an important role here as this means $g^{\mu\nu}$ is positive definite.
Doing so, the isotropic weak Kraus operator is
\begin{equation}\label{almost}
M(\{q^\alpha\}) = \exp\left(-\frac{1}{4}\Big((\boldsymbol\sigma^{-2})_{\alpha\beta}q^\alpha q^\beta
-2q^\alpha(\boldsymbol\sigma^{-2})_{\alpha\beta}\kappa^{\beta\mu}X_\mu
+\frac{n\gamma  dt}{\dim\g_o}X^2\Big)\right).
\end{equation}
Importantly, the quadratic terms are proportional to the Casimir operator
\begin{equation}\label{Casimir}
X^2 \equiv g^{\mu\nu}K(X_\mu)K(X_\nu) = (\lambda, \lambda + 2\omega)\boldsymbol{1}
\end{equation}
and therefore will not intervene with the linear generators.

Identify the coefficients of the linear generator in equation \ref{almost} with Wiener increments
\begin{equation}
dW^\mu = q^\alpha(\boldsymbol\sigma^{-2})_{\alpha\beta}\kappa^{\beta\mu}
\end{equation}
which satisfy the It\^o rule
\begin{equation}\label{ItoIto}
dW^\mu dW^\nu = g^{\mu\nu}\frac{n\gamma dt}{\dim\g_o}.
\end{equation}
If $p > \dim\g_o$, there will be many outcomes $\{q^a\}$ which result in the same displacement $dW^\mu$. One should therefore marginalize the distribution of Kraus operators to obtain the distribution of distinct Kraus operators
\begin{equation}\label{KrausBaby}
M(\{dW^\mu\}) = \exp\left(
-\frac{1}{4}\Big(\frac{\dim\g_o}{n\gamma dt}\, g_{\mu\nu}dW^\mu dW^\nu 
-2X_\mu dW^\mu
+\frac{\gamma n dt}{\dim\g_o}X^2\Big)\right)
\end{equation}
For simplicity, choose $n=\dim\g_o$
If one continues this isotropic measurement for a time $T= Nn dt$, then the final QOVM is

\begin{equation}\label{KrausTot}
\D\Z[dW] =
\exp\left(-\frac{\gamma T X^2}{4\dim\g}\right) \otimes \exp\left(-\frac{\gamma T X^{*2}}{4\dim\g}\right)
\D\mu[dW]
K[dW]\otimes K[dW]
\end{equation}
where defined is the Wiener measure
\begin{equation}\label{BigWiener}
\D\mu[dW] = \left(\frac{\gamma dt}{2\pi}\right)^{T/\dim\g_o dt} \exp\left(-\frac{1}{2}\, \int_0^T \!\! \frac{g_{\mu\nu}}{\gamma dt}dW^\mu(t) dW^\nu(t)\right)
\end{equation}
and $K[m]=K(T)$ is the solution to the differential equation
\begin{equation}\label{bestSDE}
d K(t) = \left(\frac{1}{2}X_\mu dW^\mu(t)+\frac{1}{8}X^2 \gamma dt\right) K(t)
\end{equation}
with initial condition $K(0)=1$.

\section{Discussion and Conclusion}

We have shown that the generalized-coherent-state POVM is implemented via the nonadaptive isotropic continuous measurement.
Specifically, that the ensemble of Kraus operators of the isotropic continuous QOVM are such that
\begin{equation}
\mathrm{Prob}\Big(\bar{P}\big(K^\dag K\big)< \epsilon\Big) = \mathrm{Prob}\Big(\bar{P}\big(KK^\dag\big) < \epsilon\Big) > 1 - \sqrt{\frac{2\dim\g_o}{\pi\gamma T}}\,\ln\frac{4(\omega,\lambda)}{|\lambda|^2\epsilon}.
\end{equation}
A future draft of this paper will include a survey of generalized-coherent-states as they appear in quantum information and physics.

\section*{Acknowledgements}

Thanks to Ivan Deutsch and Ezad Shojaee for originally intuiting the nonadaptive continuous isotropic measurement for spin systems.
Thanks to Carl Caves and Jonathan Gross for many productive conversations about measurement theory.
Thanks to Raf Alexander and Pablo Poggi for reading previous drafts.
Thanks to the members of CQuIC for providing such a dynamic and esteemed environment.

This work was supported by the National Science Foundation under grant PHY-1630114.

\bibliographystyle{apsrev}
\bibliography{GenCohStPOVM.bib}

\end{document}